\documentstyle[seceq]{ptptex}

\newcommand{\be}{\begin{equation}}
\newcommand{\ee}{\end{equation}}
\newcommand{\bea}{\begin{eqnarray}}
\newcommand{\eea}{\end{eqnarray}}

\def\be{\begin{equation}}
\def\ee{\end{equation}}
\def\bea{\begin{eqnarray}}
\def\eea{\end{eqnarray}}

\def\sm{M_\odot}
\def\etal{et al. }

\title{
Primordial Black Holes and Gravitational Memory
}

\author{
Bernard {\sc Carr}\footnote{E-mail address: B.J.Carr@qmw.ac.uk}
and Chad {\sc Goymer}}

\inst{
Astronomy Unit, Queen Mary \& Westfield College,
London E1 4NS, England
\\
$^*$Yukawa Institute for Theoretical Physics, 
Kyoto University, Kyoto 606-8502}

\recdate{
}

\abst{We discuss the various ways in which primordial black holes may have formed in the early Universe 
and how the effects of such black holes can be used to place constraints on cosmological models.
We show that such constraints may be severely modified if the value of the gravitational
``constant" $G$ varies with cosmological epoch, a possibility which arises in many scenarios
for the early Universe. The nature of the modification depends upon whether the value of $G$ near a black
hole maintains the value it had at its formation epoch (corresponding to gravitational memory) or
whether it tracks the background cosmological value. This is still uncertain but we discuss various
approaches which might help to resolve the issue.} 

\begin{document}

\maketitle

\section{Introduction}

It is well known that primordial black holes (PBHs) could have formed in the early Universe. \cite{rf:1,rf:2} A 
comparison of the cosmological density at any time after the Big Bang with the density associated with a black hole
shows that PBHs would have of order the particle horizon mass at their formation epoch.
PBHs could thus span an enormous mass range: those formed at the
Planck time ($10^{-43}$s) would have the Planck mass ($10^{-5}$g),
whereas those formed at 1~s would be as large as $10^5\sm$,
comparable to the mass of the holes thought to reside in galactic nuclei. 

PBHs would could arise in various ways.\cite{rf:3} Since the early Universe is unlikely to have been exactly
Friedmann,  they would form
most naturally from initial inhomogeneities but they might also form through other mechanisms at a cosmological phase
transition. We discuss these various formation mechanisms in Section
2. PBHs could also have various types of cosmological consequence. Those larger than $10^{15}$g could contribute to the
dark matter density and might even produce observable microlensing effects.
Those smaller than this would have evaporated by the Hawking mechanism \cite{rf:4} and could contribute to the
flux of cosmic rays. 

Studying the formation and cosmological consequences of PBHs is important because it enables one to
place constraints on the early Universe (eg. on the
spectrum of density fluctuations and the nature of any phase transitions). Indeed PBHs serve as a probe of times much
earlier than that associated with most other ``relicts" of the Big Bang. While photons decoupled at
$10^6$y, neutrinos at 
$1$~s and WIMPs at $10^{-10}$s, non-evaporating PBHs go back to $10^{-23}$s and evaporating ones all the way back to
the Planck time. Therefore even if PBHs never formed, their non-existence
gives interesting information. We review and update these constraints in Section 2. 

The main purpose of this paper is to examine
how the PBH constraints are modified if the value of the gravitational ``constant" $G$ was different at early times.
As reviewed in Section 3, this idea has a long history and should no longer be regarded as exotic. 
It arises in various scalar-tensor theories of gravity 
and these are a natural setting for many currently popular models of the early Universe. 
A number of astrophysical constraints suggest the $G$ could not have varied much since the epoch of cosmological 
nucleosynthesis but PBHs would have formed much earlier than this. 
Homogeneous cosmological solutions in such theories have been much studied and generally lead to a simple power-law
dependence of $G$ on
$t$. However, {\it inhomogenous} models (such as would be associated with PBH formation) would lead to a value
of $G$ varying both temporally and spatially and are much less well understood. 

Black hole formation and evaporation could be greatly
modified in variable-$G$ cosmologies, since many of their properties (eg. their  radius and Hawking temperature) 
depend explicitly on $G$. However, the nature of the modification depends upon the extent to which a PBH preserves 
the value of
$G$ at its formation epoch rather than following the background cosmological value.  Barrow \cite{rf:5} first drew
attention to this problem and introduced two possibilities: ``scenario  A'', where $G$ everywhere maintains the background
value and ``scenario B'',  where the value of $G$ stays constant near the black hole but evolves far away from it. This
last possibility he termed     ``gravitational memory''. There would be interesting modifications to the cosmological
consequences of PBH formation in both cases but they would be more dramatic in scenario B. 

Barrow \& Carr \cite{rf:6} considered the implications of these two scenarios in detail and
we will review some of their conclusions in Section 3. However, they did not address the issue of gravitational memory
itself and we turn to this in Section 4. Although the problem is still not resolved, we
will report on several interesting developments. These are described in more detail elsewhere \cite{rf:7} and exploit
two important equivalences. The first equivalence is between scalar-tensor theories and general relativity with a
scalar field. This comes about because these two theories are related by a conformal transformation and this means that the
problem of gravitational memory can be probed by investigating the formation and evolution of a black hole in a
cosmological background containing a scalar field. This leads to four variants of the gravitational memory 
scenario, which we describe in Section 5. The second
equivalence is between a scalar field and a stiff fluid (with equation of state $p=\rho$). This allows one to relate the
issue of gravitational memory to the problem of black hole accretion in a Universe containing a stiff fluid. This problem
has already been investigated by several people and, as discussed in Section 6, their findings lead to several new
insights.

\section{PBH formation and constraints on the early Universe}

One of the most important reasons for studying PBHs is that it enables one to 
place limits on the spectrum of density fluctuations in the early Universe. This
is because, if the PBHs form
directly from density perturbations, the fraction of
regions undergoing collapse at any epoch is determined 
by
the root-mean-square amplitude $\epsilon$ of the fluctuations
entering the horizon at that epoch and the equation of state 
$p=\gamma
\rho~(0< \gamma <1)$. One usually expects a radiation equation of
state ($\gamma =1/3$) in the early Universe. In order to 
collapse
against the pressure, an overdense region must be larger than the
Jeans length at maximum expansion and this is just 
$\sqrt{\gamma}$
times the horizon size. On the other hand, it cannot be larger than the horizon size, else it would
form a separate closed universe and not be part of our Universe.\cite{rf:8} 

This has two important implications. Firstly, PBHs forming at time $t$ should
have of order the horizon mass then:
\be
M(t) \approx {c^3 t\over G} \approx 10^{15}\left({t\over 10^{-
23} \ {\rm s}}\right) g.
\ee
Secondly, for a region destined to collapse to a PBH, one requires the fractional overdensity at the horizon
epoch,
$\delta$, to exceed $\gamma$. Providing the desnity 
fluctuations have a Gaussian distribution and are spherically
symmetric, one can infer that the fraction of regions of mass 
$M$ which collapse is \cite{rf:9}
   \be
\beta(M) \sim \epsilon(M) \exp\left[-{\gamma^2\over 
2\epsilon(M)^2}\right]
   \ee
where $\epsilon (M)$ is the value of $\epsilon$ when the 
horizon
mass is $M$. The PBHs can have an extended mass spectrum only 
if the
fluctuations are scale-invariant (i.e. with $\epsilon$ 
independent of $M$) but this is expected in many scenarios.
Recent hydrodynamical calculations for the $\gamma=1/3$ case have refined these estimates somewhat:   
Niemeyer \& Jedamzik \cite{rf:10} find that one needs $\delta >0.8$ rather than $\delta >0.3$ to ensure PBH 
formation, and Shibata \& Sasaki \cite{rf:11} reach similar conclusions, but the same basic picture still applies. 

Another interesting development has been the application of ``critical phenomena"
to PBH formation.  Studies of the collapse of various types of spherically symmetric matter fields have shown that there
is  always a critical solution which separates those configurations which form a black hole from 
those which disperse to an asymptotically flat state. The configurations are described by some index $p$ and, as the
critical index $p_c$ is approached, the black hole mass is found to scale as $(p-p_c)^{\eta}$ for some exponent
$\eta$. This effect was first discovered for scalar fields \cite{rf:12} but subsequently
demonstrated for radiation \cite{rf:13} and then more general fluids with equation of state $p=\gamma \rho$.
\cite{rf:14,rf:15} 

In all these studies the spacetime was assumed to be asymptotically flat. However, Niemeyer \& Jedamzik \cite{rf:16} have 
recently applied
the same idea to study black hole formation in asymptotically Friedmann models and have found similar results. 
For a variety of initial density perturbation profiles, they find that the
relationship between the PBH mass and the the horizon-scale density perturbation has the form
\be
M = K M_H(\delta - \delta_c)^{\gamma}
\ee
where $M_H$ is the horizon mass and the constants are in the range $0.34<\gamma<0.37$, $2.4<K<11.9$ and
$0.67<\delta_c <0.71$ for the various configurations. Since $M \rightarrow 0$ as $\delta \rightarrow \delta_c$, 
this
suggests that PBHs may be much smaller than the particle horizon at formation (although it is clear that a fluid
description must break down if they are too small) and it also modifies the mass spectrum. \cite{rf:17,rf:18}

These developments might be regarded as refinements of the original scenario since they all presuppose 
a relationship of the form (2.2). However, in some situations
eqn (2.2) would fail qualitatively. For example,  PBHs would form more easily if the
equation of state of the Universe were ever soft ($\gamma\ll$1). This might apply if there was a phase 
transition which channelled the mass of the Universe into non-relativistic 
particles or which temporally reduced the pressure. In this case, only those regions which are sufficiently spherically
symmetric at maximum expansion can undergo collapse; the dependence of $\beta$ on $\epsilon$ would then be
weaker than indicated by eqn (2.2) but there would still be a
unique relationship between the two parameters. \cite{rf:19}

The fluctuations required to make the PBHs may either be primordial or they may arise spontaneously at some epoch. 
One natural source of fluctuations would be inflation \cite{rf:20,rf:21} and, in this
context, $\epsilon(M)$ depends implicitly on the inflationary potential. Indeed many people have studied PBH formation
in inflationary scenarios as an important way of constraining this potential.  
\cite{rf:22}\tocite{rf:28} Recently the Gaussian assumption has been questioned in the inflationary
context,\cite{rf:29,rf:30} so eqn (2.2) may not apply, but one still finds that
$\beta$ depends very sensitively on $\epsilon$.

Some formation mechanisms for
PBHs do not depend on having primordial
fluctuations at all.  For example,  at any spontaneously broken symmetry epoch, PBHs might form 
through the collisions of bubbles of broken symmetry.\cite{rf:31}\tocite{rf:33} 
PBHs might also form spontaneously through the
collapse of cosmic strings.\cite{rf:34}\tocite{rf:38} In these
cases $\beta(M)$  depends, not on $\epsilon(M)$, but
on other cosmological parameters, such the bubble formation
rate or the string mass-per-length. 

In all these scenarios, the current density parameter $\Omega_{PBH}$ associated
with PBHs which form at a redshift $z$ or time $t$ is related to
$\beta$ by \cite{rf:9}
   \be
\Omega_{\rm PBH} = \beta\Omega_R(1+z) \approx 10^6 
\beta\left({t\over s}\right)^{-1/2} \approx 10^{18}\beta\left({M\over 10^{15}g}\right)^{-1/2}
   \ee
where $\Omega_R \approx 10^{-4}$ is the density parameter of the microwave
background and we have used eqn (2.1). The $(1+z)$ factor arises 
because the radiation density scales as $(1+z)^4$, whereas the PBH
density scales as $(1+z)^3$. Any limit on $\Omega_{PBH}$ therefore places a constraint on $\beta(M)$ and 
the constraints are summarized in Fig. 1.
The constraint for non-evaporating mass ranges above
$10^{15}$g comes from requiring
$\Omega_{PBH}<1$.
Much stronger constraints are associated with PBHs
which were smaller than this, since they would have evaporated by now. For example, the 
constraints below $10^{6}$g are based on the (not necessarily secure) assumption that evaporating PBHs leave stable Planck
mass relics, in which case these  relics are required to have less than the critical density. \cite{rf:23,rf:39,rf:40}
Other constraints are associated with the generation of entropy, modifications to the cosmological production of light 
elements and the contribution to the cosmological $\gamma$-ray background.

The constraints in Fig. 1 are discussed in detail by Carr \etal \cite{rf:23} but we note that Kohri \& Yokoyama
\cite{rf:41} have recently improved the constraints on $\beta(10^8 - 10^{10}g)$ which come from cosmological
nucleosynthesis considerations. Here we wish to emphasize
that the strongest constraint is the cosmic ray limit associated with the $10^{15}$g PBHs evaporating
at the present epoch.\cite{rf:42}\tocite{rf:46} Indeed the recent detection of a Galactic $\gamma$-ray background
\cite{rf:47},  measurements of the antiproton flux,\cite{rf:48} and the discovery of very short period
$\gamma$-ray  burts,\cite{rf:49} may even provide positive evidence for
such PBHs. This is discussed in detail elsewhere. \cite{rf:50} The constraints on 
$\beta(M)$ can be converted into constraints on 
$\epsilon(M)$ using eqn (2.2) and these are shown in Fig. 2. Also shown here are the (non-PBH) constraints
associated with the spectral distortions in the cosmic microwave background induced by the dissipation of 
intermediate scale density perturbations and the COBE quadrupole measurement, as well as lines corresponding
to various slopes in the $\epsilon (M)$ relationship.    

Finally it should be emphasized that there has been particular interest recently in whether PBHs could have 
formed at the quark-hadron phase transition at $10^{-5}$s. This is because the horizon mass is of order $1\sm$ then, so
such PBHs would naturally have the sort of mass required to explain the MACHO microlensing results.\cite{rf:51}  
One might expect PBHs to form more easily at that epoch because of a temporary softening of the equation of 
state. If the QCD phase 
transition is assumed to be of 1st order, then hydrodynamical calculations show that the value of
$\delta$ required for PBH formation is indeed reduced below the value which pertains in the radiation case.\cite{rf:52} 
This means that 
PBH formation will be strongly enhanced at the QCD epoch, with the mass distribution being peaked around the
horizon mass then.\cite{rf:18} 

Since the focus of this meeting is gravitational waves, it should be emphasized that one of the interesting
implications of the PBH MACHO scenario is the possible existence of a halo population of {\it binary} black
holes.\cite{rf:53} In this case, there could be $10^8$ binaries inside 50 kpc and some of these could be coalescing due to
gravitational radiation losses at the present epoch.\cite{rf:54} Current interferometers
(LIGO, VIRGO, GEO, TAMA) could detect such coalescences within 50 Mpc, corresponding to a few events per year. Future
space-borne interferometers (such as LISA or OMEGA) might detect 100 coalescences per year. If the associated gravitational
waves were detected, it would provide a unique probe of the halo distribution (eg. its density
profile and core radius).\cite{rf:55}

\begin{figure}\label{F1}
\vspace{3.0in}
\caption{Constraints on $\beta(M)$}
\end{figure}

\begin{figure}\label{F2}
\vspace{3.0in}
\caption{Constraints on $\epsilon(M)$}
\end{figure}

\section{Cosmology in varying-G theories}

Most variable-G scenarios associate the gravitational ``constant'' with some form of scalar field
$\phi$. This notion has its roots in Kaluza-Klein theory, in which a scalar field appears
in the metric component $g_{55}$ associated with the 5th dimension. 
Einstein-Maxwell theory then requires that this field be related to $G$. \cite{rf:56} Although this was assumed constant
in the original Kaluza-Klein theory,
Dirac \cite{rf:57} noted the the ratio of the electric to gravitational force between to protons ($e^2/Gm_p^2$)
and the ratio of the age of the Universe to the atomic timescale ($t/t_a$) and the square-root of the number of
particles in the Universe ($\sqrt{M/m_p}$) are all comparable and of order $10^{40}$. This unlikely coincidence led him
to propose that these relationships must {\it always} apply, which requires 
\be
G \propto t^{-1}, \;\;\;GM/R \sim 1,
\ee
where $R\sim ct$ is the horizon scale. The first condition led Jordan \cite{rf:58} to propose a theory in which the scalar
field in Kaluza-Klein theory is a  function of {\it both} space and time, and this then implies that $G\sim \phi^{-1}$ has
the same property. The second condition
implies the Mach-type relationship
$\phi \sim M/R$, which suggests \cite{rf:59} that $\phi$ is a solution of the wave equation $\Box \phi \sim \rho$. This
motivated Brans-Dicke (BD) theory,\cite{rf:60} in which the Einstein-Hilbert Lagrangian is replaced by
\be
L = \phi R - \frac{\omega}{\phi}\phi_{,\mu}\phi_{,\nu}g^{\mu \nu} + L_m \;,
\ee
where $L_m$ is the matter Langrangian and the constant $\omega$ is the BD parameter. The potential $\phi$
then satisfies 
\be
\Box \phi = \left(\frac{8\pi}{2\omega +3}\right) T,
\ee
where $T$ is the trace of the matter stress-energy tensor, and this has the required Machian form. Since $\phi$ must
have a contribution from {\it local} sources of the form
$\Sigma_i (m_i/r_i)$, this entails a violation of the Strong Equivalence Principle. In order to test this, the PPN
formalism was introduced. Applications of this test in a variety of astrophysical situations (involving the
solar system, the binary pulsar and white dwarf cooling) currently require
$|\omega| > 500$, which implies that the deviations from general relativity can only ever be small in BD
theory.\cite{rf:61}

The introduction of generalized scalar-tensor theories, \cite{rf:62}\tocite{rf:64} in which $\omega$ is itself a function
of $\phi$, led to a considerably broader range of variable-$G$ theories. In  particular, it permitted the possibility that
$\omega$ may have been small at early times (allowing noticeable variations of $G$ then) even if it is large
today. In the last decade interest in such theories has been revitalized as a result of early Universe studies.
Inflation theory \cite{rf:65} has made the introduction of scalar fields almost mandatory and
extended inflation specifically requires a model in which $G$ varies.\cite{rf:33}
In higher dimensional Kaluza-Klein-type cosmologies, the variation in the sizes of the extra dimensions also naturally
leads to a variation in $G$. \cite{rf:66}\tocite{rf:68}. The currently popular
low energy string cosmologies necessarily involve a scalar (dilaton) field \cite{rf:69} and bosonic superstring
theory, in particular, leads \cite{rf:70} to a Lagrangian of the form (3.2) with $\omega =-1$.

The intimate connection between dilatons, inflatons and scalar-tensor theory arises because one can always transform
from the (physical) Jordan frame to the Einstein frame, in which the Lagrangian has the standard Einstein-Hilbert form
\cite{rf:71}
\be
L = \bar{R} - 2\psi_{,\mu}\psi_{,\nu}\bar{g}^{\mu \nu} + L_m.
\ee
Here the new scalar field $\psi$ is defined by 
\be
\mathrm{d}\psi = \left(\frac{2\omega+3}{2}\right)^{1/2}\frac{\mathrm{d}\phi}{\phi}
\ee
and the barred (Einstein) metric and gravitational constant are related to the
original (Jordan) ones by 
\be
g_{\mu \nu} = A(\phi)^2 \bar{g}_{\mu \nu},\;\;\;G = [1+\alpha^2(\phi)] A(\phi)^2 \bar{G}, \quad \alpha \equiv A'/A,
\ee
where the function $A(\phi)$ specifies a conformal transformation. Thus scalar-tensor theory can be interpreted as
general relativity plus a scalar field.

The behaviour of homogeneous cosmological models in BD theory is well understood.\cite{rf:72} Their crucial feature is
that they are vacuum-dominated at early times but always tend towards the general relativistic solution during the
radiation-dominated era. This is a consequence of the fact that the radiation energy-momentum tensor is trace-free
[i.e. $T=0$ in eqn (3.3)]. This
means that the full radiation solution can be approximated by joining a
BD vacuum solution to a general relativistic radiation solution at some time $t_1$, which 
may be regarded as a free parameter of the theory. However, when the matter density becomes greater than the radiation
density at $t_e \sim 10^{11}$s,  
the equation of state becomes that of dust $(p = 0)$ and $G$ begins to vary again. For a $k = 0$ model, one can show 
that in the three eras \cite{rf:6}
\be
G = G_0 (t_0/t_e)^n, \quad a \propto t^{(2-n)/3} \quad (t>t_e)
\ee
\be
G = G_e \equiv G_0 (t_0/t_e)^n , \quad  a \propto t^{1/2} \quad (t_1<t<t_e)
\ee
\be
G = G_e (t/t_1)^{-(n + \sqrt{4n + n^2})/2}, \quad a \propto t^{(2 - n - \sqrt{4n + n^2})/6} \quad (t<t_1)
\ee
where $G_0$ is the value of $G$ at the current time $t_0$, $n \equiv 2/(4 + 3\omega)$ and $(t_0 / t_e) \approx 10^6$. 

Since the BD coupling constant is constrained by $|\omega| > 500$, which implies 
$|n| < 0.001$, eqns (3.7) to (3.9) imply that the deviations from general relativity are never large if the value of
$n$ is always the same. However, for reasons explained below, it is also interesting to consider BD models
in which $n$ and $\omega$ can violate the current constraints. This allows considerably more exotic behaviour,
especially in the vacuum-dominated era. In particular, we note the following
features:\cite{rf:6} models with $\omega <-3/2$ are probably excluded because the energy density of the scalar field is
negative; models with
$-3/2<\omega<-4/3$ undergo power-law inflation during the vacuum-dominated era because the exponent of
$t$ in the expression for $a$ in eqn (3.9) exceeds $1$; models with $-4/3<\omega<0$ can bounce during the
vacuum-dominated era because this exponent is negative.

The behaviour of cosmological models in more general scalar-tensor theories depends on the form of $\omega(\phi)$
but they still retain the feature that the general relativistic solution is a late-time attractor during the radiation
era.  Since one requires $G\approx G_0$ to $10\%$ at the epoch of primordial nucleosynthesis, \cite{rf:72} one needs the
vacuum-dominated phase to end at some time $t_v <$ 1 s. The theory approaches general relativity in the weak field limit
only if
$\omega \rightarrow \infty$ and $\omega '/\omega ^3 \rightarrow 0$ (where a prime denotes $d/d\phi$) but $\omega(\phi)$ is
otherwise unconstrained. Barrow \& Carr consider a toy model in which
\be
2\omega +3 = 2\beta (1-\phi/\phi_c)^{-\alpha}
\ee
where $\alpha$ and $\beta$ are constants. This leads to
\be
2\omega +3 \propto t^{-\alpha/(2-\alpha)}, \quad \omega '/\omega ^3 \propto t^{(1-2\alpha)/(1-\alpha)} \quad (t<t_v),
\ee
so one requires $1/2<\alpha<2$ in order to have $\omega \rightarrow \infty$ and $\omega '/\omega ^3
\rightarrow 0$ as $t \rightarrow \infty$. In the $\alpha =1$ case, one finds
\be
G \propto t^{-2\lambda/(3-\lambda)}, \quad a \propto t^{(1-\lambda)/(3-\lambda)}, \quad \lambda \equiv
\sqrt{3/(2\beta)} \quad \;\;\; (t<t_v).
\ee
During the vacuum-dominated era, such models can therefore be regarded 
as BD
solutions in which $\omega$ is determined by the parameter $\beta$ and unconstrained by any limits on $\omega$ at the
present epoch. After $t_v$, $G$ is constant and one has the standard radiation-dominated or dust-dominated  
behaviour. 

On the assumption that a PBH of mass $M$ has a temperature and mass-loss rate \cite{rf:4}
\be
T = (8\pi GM)^{-1},\quad \dot{M} \approx - (GM)^{-2},
\ee
with $G$ having the value $G(t)$ in scenario A and $G(M)$ in scenario B, Barrow \& Carr calculate the evaporation 
time $\tau$ for various values of the parameters $n$ and $t_1$ in BD theory. The results are shown in Fig. 3(a) for
scenario A and Fig. 3(b) for scenario B. Here $M_*$ is the mass of a PBH evaporating at the present
epoch, $M_e$ is the mass of a PBH evaporating at time $t_e$ and $M_{crit}$ is the mass of a PBH
evaporating at the present epoch in the standard (constant $G$) scenario. In scenario A with $n<-1/2$, there is
a maximum mass of a PBH which can ever evaporate and this is denoted by $M_{\infty}$. The results for the
scalar-tensor with
$\omega(\phi)$ given by eqn (3.10) with
$\alpha=1$ are  shown in Fig. 3(c) for scenario B with various values of the parameters $\lambda$ and $t_v$.  
The corresponding modifications to the constraints on $\beta(M)$ in all three cases are shown in Fig. 3(d), which should be
compared to Fig. 1. 

\begin{figure}\label{F3}
\vspace{4.3in}
\caption{Dependence of the PBH evaporation time $\tau$ on initial mass $M$ in (a) BD theory with scenario A, (b) 
BD theory with scenario B, (c) scalar-tensor theory with $\alpha=1$ and scenario B. Also shown are (d) the modifications to
the constraints on $\beta(M)$ in these cases.}
\end{figure}

\section{Black holes in scalar-tensor theory} 

In BD theory Hawking \cite{rf:73} showed that, providing the weak energy condition holds, the gradient 
of $\phi$ must be zero everywhere for stationary, asymptotically flat black holes. This means that such black 
holes are identical to those in general relativity.  This result can be generalized
to all scalar-tensor theories and suggests that such theories are in agreement with the ``no-hair'' theorem.

Numerical calculations support this theorem. \cite{rf:74}\tocite{rf:76}
An Oppenhiemer-Snyder type collapse reveals outgoing scalar gravitational radiation, which 
radiates away the scalar mass. When all
the scalar mass is lost, the black hole settles down with a constant scalar field to the Schwarzschild form. The scalar
radiation generated in this way would be of great interest in itself. As shown by  Harada et al.,\cite{rf:76}
if a black hole of mass
$M$ collapses, the ratio of the amplitudes of the scalar and tensor gravitational waves is
\be
\frac{h_S}{h_T} \sim 10^{-3} \left(\frac{500}{f}\right)\left(\frac{10GM}{a}\right)^{-1}\frac{e^2}{\sqrt{1-e^2}},
\quad
f \sim \left(\frac{M}{\sm}\right)^{1/2}\left(\frac{r_s}{15km}\right),
\ee
where $a$ is the initial size of the region, $e$ is its eccentricity and $f$ is the frequency. There might also be a
stochastic background of scalar gravitational waves generated by the formation of PBHs in the early Universe, although its
density would have been diluted by now because of redshift effects.

It should be pointed out that the simulations of scalar collapse also show a rather unusual
feature:\cite{rf:74,rf:75} the area of the event horizon decreases for a time, violating the
area theorem, and it may also lie inside the apparent horizon during this period. This occurs for all 
values of $\omega$ and is due to the fact that the scalar 
field violates the weak energy condition. This condition holds in the Einstein 
frame as long as $\omega > -3/2$ but it need not be true in the physical Jordan frame. Examples have 
been found \cite{rf:77}\tocite{rf:79} of black holes in BD theory which have scalar hair but these
arise only for certain ranges of $\omega$: $\omega < -4/3$ in \cite{rf:77}, $\omega < -3/2$ in \cite{rf:78} and 
$-5/2 \leq \omega < -3/2$ in \cite{rf:79}. It is unclear that these can represent the end state of realistic
gravitational collapse.

In addressing the question of which of Barrow's scenarios A or B is most plausible, it should be stressed that 
the scalar
no hair theorem has only been proved for asymptotically flat spacetimes, so it is not clear that it also applies in the
asymptotically Friedmann case. While the no hair theorem suggests that $\phi$ should tend to a {\it locally} 
constant value (close to the black hole), it is not obvious that this needs to be the asymptotic cosmological value.
Indeed, since the homogeneizing of $\phi$ is only ensured by scalar wave emission, one might infer that this can only be
achieved on scales less than the particle horizon. 

The only way to determine what happens is to seek a precise mathematical model
for a  black hole in a cosmological background.  
One approach is to try matching the black hole and cosmological solutions over some boundary
$\Sigma$. An example of such a matching in general  relativity is the Einstein-Straus or ``Swiss cheese"
model. \cite{rf:80} Here a Friedmann exterior is matched with a general spherically symmetric interior.
If there is no scalar field, it turns out that the latter has to be the static Schwarzschild solution but the situation may
be  more complicated in the present context due to the presence of scalar gravitational radiation. 
In general one can show that the following continuity conditions must apply at $\Sigma$:
\be
[g_{\mu \nu}] = 0, \quad [G_{\mu \nu} n^{\mu} n^{\nu}] = [G_{\mu \nu} u^{\mu} n^{\nu}] = 0, 
\quad [\phi] = 0, \quad [\phi_{\mu}n^{\mu} ] = 0
\ee
where $n^{\mu}$ and $u^{\mu}$ are 4-vectors normal and tangent to $\Sigma$, respectively. \cite{rf:81}

Unfortunately, it turns out that an Einstein-Straus type 
solution does not exist in BD theory. This is because the only way to satisfy the 
junction conditions (4.2) is if $\phi$ is spatially and temporally constant, which is just 
the general relativistic case. However, Oppenheimer-Snyder collapse has been investigated 
\cite{rf:76,rf:82} in which a ball of dust described by a $k=+1$ Friedmann interior and a Schwarzschild exterior 
collapses to a black hole. In these calculations the scalar field is taken to be constant before the 
collapse and its back-reaction on the metric is assumed to be always negligible. It is found that, as 
the collapse proceeds, a scalar gravitational wave propagates outwards before the scalar 
field settles down to being constant again. In principle, it should be possible to extend this
approach to the present problem by attaching a $k=+1$ Friedman interior to a $k=0$ Friedman exterior.

Jacobsen \cite{rf:83} has addressed the problem analytically by looking for a spherically symmetric solution which
represents a black hole in an asymptotically Friedmann model in which $\phi$ satisfies
the wave equation
\be
\left(-g^{tt}\partial_t^2 + \frac{1}{\sqrt{-g}} \partial_r [\sqrt{-g}\; g^{rr}\partial_r ]\right) \phi(t,r) =0.
\ee
As $r \rightarrow \infty$, he assumes that $\phi$ asymptotes to
\be
\phi_c(t) \approx \dot{\phi_c}t
\ee
in the Einstein frame. This presupposes that the black hole event horizon is much smaller than the particle
horizon, so that the cosmological timescale is much longer than the black hole timescale, in which case $\dot{\phi_c}$
in eqn (4.4) can be regarded as constant. He then seeks a solution which is a combination of a homogeneous 
part $\phi_1(t) $ and a time-independent part $\phi_2(r)$, where
\be
\phi_1 = \dot{\phi_c}t, \quad \phi_2 = \ln(1-r_H/r).
\ee
Both these terms diverge at the event horizon ($r_H$) but one can find a combination of them which is regular there:
\be
\phi_3 = \phi_1 + r_o \phi_2 \dot{\phi_c} =  \dot{\phi_c}[v-r-r_Hln(r/r_H)]
\ee 
where $v$ is advanced ``tortoise" time, together with a superposition of waves falling into the black hole or dispersing
to infinity. The equipotential $\phi =\phi_3$ intersects the event horizon at $v=v_H$ and infinity at $t=t_{\infty}$. Since
$v=v_H$ and  and $t=t_{\infty}$ themselves intersect at $r\approx 1.5GM$, he infers that there is little lag between the
value of $\phi$ at the event horizon and particle horizon, which means that the memory can only be weak. 

Although Jacobsen's analysis demonstrates that a solution with little memory does exist, it must be emphasized 
that he has really put this feature into the solution at the outset by assuming the black hole is much smaller
than the particle horizon. This assumption is inappropriate if the
black hole has a size {\it comparable} to the particle horizon at formation (as applies for a PBH), so 
the field around a collapsing region need not necessarily evolve to the form (4.6).  Another rather
crucial feature of his analysis is that he claims that, while the PBH mass is nearly constant in the Einstein frame, it
increases as
\be
M = A(\phi)^{-1} \bar{M} \propto \phi^{1/2}
\ee
in the Jordan frame. If this were correct, it would invalidate the analysis of Barrow \& Carr and hence Figs (3).
However, we would argue that it is not clear in which frame the mass should be taken to be constant.

\section{Variations of gravitational memory}

We now consider the question of what happens to black holes in BD or more general scalar-tensor theories during the
evolution of the Universe. In general the present value of the scalar field, $\phi(t_0)$, will be different from its
value when the black hole was formed, 
$\phi(t_f)$. In the following discussion, we will characterize the degree of gravitational memory 
by comparing the value of the scalar field at the black hole event horizon ($\phi_{EH}$) and the cosmological 
particle horizon ($\phi_{PH}$). We first
consider the two extreme situations described by Barrow \cite{rf:5}:

\be
Scenario\;A: \quad \quad \phi_{EH}(t) = \phi_{PH}(t) \quad \mathrm{for\ all}\ t
\ee
A Schwarzschild black hole forms at time $t_f$ with its event horizon radius being 
$R_f = 2G(t_f)M$. If $G(t)$ evolves with time, then the black hole adjusts quasi-statically 
through a sequence of Schwarzschild states approximated by $R = 2G(t)M$, see Fig. 4(a). 
In this scenario there are no stationary black holes when 
$G(t)$ is changing and no gravitational memory.

\be
Scenario\; B: \quad \quad  \phi_{EH}(t) = \phi_{EH}(t_f) \quad \mathrm{for\ all}\ t
\ee
A Schwarzschild black hole of size $R_f$ forms at time $t_f$ and, while $G(t)$ equals
the evolving background value beyond some scale-length 
$R_m \geq R_f$, it remains constant within $R_m$, see Fig. 4(b). In this case the black hole size
is determined by $G(t_f)$ even at the present epoch and this means that the 
region $R < R_m$ has a memory of the gravitational ``constant'' at the 
time of its formation.
\vskip .2in

Neither of these scenarios can be completely realistic since they both assume that $\phi$ is 
homogeneous almost everywhere.  However, even if $\phi$ were homogeneous initially,  
one would expect it to become inhomogeneous as collapse proceeds. Indeed, if the background value is 
increasing (as usually applies), one would expect
$\phi$ in the collapsing region to become first {\it larger} than the background value on a local dynamical 
timescale and then {\it smaller} than it on a cosmological timescale. Such behaviour would
necessarily entail a variation of $\phi$ in space as well as time. 
Solutions have been found  for specific $\omega$ in which $\phi$ is inhomogeneous \cite{rf:84}, so such models
are clearly possible in principle. We must also
allow for the possibility that
$\phi$ may vary interior to $R_m$ but on a slower or faster timescale than the background. We therefore propose two
further scenarios:
\vskip .2in

\be
Scenario\;C: \quad \quad  |\dot{\phi}_{EH}(t)| \geq |\dot{\phi}_{PH}(t)| \quad \mathrm{for\ all}\ t
\ee
where the dot represents a time derivative. This 
implies that the scalar field evolves faster at the event horizon than at the 
particle horizon until it eventually becomes homogeneous, see Fig. 4(c). We describe this as
\emph{short-term} gravitational memory and it reduces to scenario A as the timescale to become homogeneous tends to zero.
This would apply, for example,
if $\phi$ were to change on the dynamical timescales of the black hole since this is usually less than
that of the cosmological background.

\be
Scenario \;D: \quad \quad  \label{beta}
|\dot{\phi}_{EH}(t)| < |\dot{\phi}_{PH}(t)| \quad \mathrm{for\ all}\ t
\ee
This implies that $\phi$ evolves faster at the particle horizon than 
the event horizon, see Fig. 4(d). We describe this is as \emph{weak} gravitational memory 
and it reduces to scenario B 
when the left-hand-side of eqn (\ref{beta}) is zero. In this case, the evolution of $\phi$ is  
again dominated by the black hole inside some length-scale $R_m$. 
Note that, in either this scenario or the last one, the length-scale $R_m$ need not be fixed, since it could either grow or
shrink as  scalar gravitational radiation propagates. A particular example of this, to which we return shortly, would be 
\emph{self-similar} gravitational memory, in which the ratio of $\phi_{EH}$ to $\phi_{PH}$ always remains the same.  
\vskip .2in

\begin{figure}\label{F4}
\vspace{4.0in}
\caption{Different possiible forms for the evolution of the scalar field profile $\phi(r)$ for
(a) no memory, (b) strong memory, (c) short-term memory, (d) weak memory.}
\end{figure}

If $\phi$, or equivalently $G$, is inhomogeneous, then one has to consider carefully what is meant by the Hawking
temperature. In general relativity the temperature of a Schwarzschild black hole is given by eqn (3.13).
However, in a scalar-tensor theory $G \equiv G(t, r)$, so one must decide which value of $G$ is appropriate.
Since the particle creation responsible for the effect takes place near the event horizon, perhaps the local value 
$G_{EH}$ should be used, giving
\be
\label{TempEH}
T_H = (8\pi G_{EH} M)^{-1}.
\ee
On the other hand, since the radiation is only measured asymptotically, one might argue that the asymptotic
value $G_{\infty}$ is appropriate, giving 
\be
\label{TempInf}
T_H = (8\pi G_{\infty} M)^{-1}.
\ee
It is not certain whether eqn (\ref{TempEH}) or (\ref{TempInf}) should be used. If eqn (\ref{TempEH}) applies, then the
temperature depends crucially on whether or not gravitational memory exists. However, if eqn (\ref{TempInf}) applies, the
temperature is unaffected by this and just depends on the background.

Gravitational memory has also been investigated in the context of boson stars \cite{rf:85}. 
A boson star is the analogue of a neutron star, formed when a large collection of bosonic 
particles become gravitationally bound \cite{rf:86,rf:87}. In this case, weak gravitational memory is 
already known to occur. An interesting feature of this is that two boson stars of the 
same mass may differ in other physical properties (eg. radius), depending on when they 
formed. However, the situation may be very different for a black hole since this has an event horizon.

\section{Gravitational memory and the accretion of a stiff fluid}

In general relativity there is an equivalence between a scalar field and a stiff 
fluid and we now show how this can be exploited in studying gravitational memory. In the Einstein frame, the energy
momentum tensor for a perfect fluid is
\be
\bar{T}_{\mu \nu} = (\rho + p)u_{\mu} u_{\nu} + \bar{g}_{{\mu \nu}} p
\ee
where $u_{\mu}$ is the velocity of the 
fluid. If we define a velocity field by
\be
\label{FVel}
u_{\mu} = \frac{\bar{\phi}_{\mu}}{(-\bar{g}^{\rho \sigma} \bar{\phi}_{\rho} \bar{\phi}_{\sigma})^{1/2}}\;,
\ee
this gives
\be
\bar{T}_{\mu \nu} = -\frac{(\rho + p) \bar{\phi}_{\mu} \bar{\phi}_{\nu}}{\bar{g}^{\rho \sigma} \bar{\phi}_{\rho}
\bar{\phi}_{\sigma}} + p \bar{g}_{\mu \nu}\;.
\ee
By comparing this to the energy-momentum tensor for a scalar field, we find that
\be
p = \rho = -\frac{1}{2} \bar{g}^{\rho \sigma} \bar{\phi}_{\rho} \bar{\phi}_{\sigma},
\ee
so we have a stiff fluid. This 
equivalence applies provided that the derivative of the scalar field is timelike. Otherwise the 
velocity field defined in (\ref{FVel}) would be imaginary. 

This is relevant to the gravitational memory problem because we can now interpret the various scenarios 
discussed in Section 5 in terms of the {\it accretion} of a stiff fluid. 
If the
black hole does not accrete at all or accretes very little, this will correspond to 
strong or weak gravitational memory (scenarios B and D, respectively). However, if enough accretion occurs to homogenize
$\phi$, this will correspond to short-term gravitational memory (scenario C). The faster the accretion, the shorter the
memory, so scenario A corresponds to the idealization in which homogenization is instantaneous.

A simple Newtonian treatment \cite{rf:1} for a general fluid suggests that the accretion rate in the Einstein 
frame should be
\be
\dot{M} = 4\pi \rho R^2_A v_s,
\ee
where $R_A= G M/ v_s^2$ is the accretion radius and $v_s$ is the sound-speed in the accreted fluid. For a stiff fluid,
$v_s=c$ and $R_A = G M/c^2$, while $\rho \sim 1/(Gt^2)$ in a Friedmann universe at early times, so we have 
\be
\frac{\mathrm{d}M}{\mathrm{d}t} \approx \frac{G M^2}{c^3 t^2}.
\ee
This can be integrated to give
\be
\label{mass}
M \approx \frac{c^3t/G}{1 + \frac{t}{t_f} \left(\frac{t_f}{M_f} - 1 \right)}
\ee
where $M_f$ is the black hole mass at the time $t_f$ when it formed. If we assume that $M_f = \eta t_f$ with $\eta <1$,
then eqn (\ref{mass}) implies
\be
M\rightarrow M_f(1 - \eta)^{-1} \quad \mathrm{as} \quad t\rightarrow \infty.
\ee
If $\eta \ll 1$, the black hole could not grow very much. However, if $\eta$ is close to 
1, which must be the case if $v_s \approx c$, then the black hole could grow significantly. 
In particular, in the limit $\eta =1$, eqn (\ref{mass}) implies $M \sim t$, so the black hole grows 
at the same rate as the universe. This simple calculation suggests that a black hole surrounded by a 
stiff fluid can accrete enough to grow at the same rate as the Universe.

Since the above calculation neglects the effects of the cosmological expansion, one needs a relativistic
calculation to check this. The Newtonian result suggests that one should look for a spherically symmetric
{\it self-similar} solution, in which every dimensionless variable is a function of $z=r/t$, so that it is
unchanged by the transformation 
$t \rightarrow at,\ r \rightarrow ar$ for any constant $a$. This problem has an interesting but rather
convolved history. By looking for a black hole solution attached to an exact Friedmann solution via a sonic point, Carr \&
Hawking first showed that there is no such solution for a radiation fluid \cite{rf:8} and the argument can be 
extended to a general $p=\gamma \rho$ fluid with
$0<\gamma <1$.  Lin et al. \cite{rf:87} subsequently claimed that there is such a solution in the special case $\gamma =1$.
However, Bicknell
\& Henriksen \cite{rf:88} then showed that this solution is unphysical, in that the density gradient diverges at the 
event horizon. The solution can be completed only by attaching a Vaidya ingoing radiation solution
interior to some surface (i.e. the scalar field has to turn into a null fluid). If one regards this as unphysical, this
suggests that the black hole must soon become much smaller than the particle horizon, after which 
eqn (6.7) implies there will be very little further accretion. Therefore the stiff fluid analysis suggests that there
should be at least \emph{weak} gravitational memory.

\section*{Acknowledgements}
The authors thank the Yukawa Institute for Theoretical Physics at Kyoto University for hospitality received during
the preparation of this paper. BJC also thanks Noriko Komatsu for patiently 
encouraging its completion and for help with the style-file.


\begin{thebibliography}{99}

\bibitem{rf:1}Ya.B. Zeldovich and I.D. Novikov,  Sov. Astron. A.J. 10 (1967), 602.
\bibitem{rf:2}S.W. Hawking, MNRAS 152 (1971), 75.
\bibitem{rf:3}B.J. Carr, in {\it Observational and Theoretical Aspects of Relativistic
Astrophysics and Cosmology}, ed. J.L.Sanz \& L.J.Goicoechea (World Scientific, 1985) p 1.
\bibitem{rf:4}S.W. Hawking, Nature 248 (1974), 30.
\bibitem{rf:5} J. D. Barrow, Phys. Rev. D. 46 (1992), R3227. 
\bibitem{rf:6} J. D. Barrow and B. J. Carr, Phys. Rev. D. 54 (1996), 3920. 
\bibitem{rf:7} C. Goymer and B. J. Carr, preprint (1999).
\bibitem{rf:8} B. J. Carr and S. W. Hawking, MNRAS 168 (1974), 399. 
\bibitem{rf:9} B.J. Carr, Ap.J. 201 (1975), 1.
\bibitem{rf:10} J. Niemeyer and K. Jedamzik, Phys.Rev.D. 59 (1999), 124013. 
\bibitem{rf:11}M. Shibata and M. Sasaki, Phys.Rev.D. (1999) (in press); gr-qc/9905064.
\bibitem{rf:12}M.W. Choptuik, Phys. Rev. Lett. 70 (1993), 9.
\bibitem{rf:13}C. R. Evans and J. S. Coleman, Phys. Rev. Lett. 72 (1994), 1782.
\bibitem{rf:14}D. Maison, Phys. Lett B 366 (1996), 82.
\bibitem{rf:15}T. Koike, T. Hara and S. Adachi, Phys. Rev. D 59 (1999), 104008.
\bibitem{rf:16} J. Niemeyer and K. Jedamzik, Phys.Rev.Lett. 80 (1998), 5481.  
\bibitem{rf:17}H. Kim, preprint (1999); poster presentation.
\bibitem{rf:18} A.M. Green and A.R. Liddle, Phys. Rev. D. 60 (1999), 063509.
\bibitem{rf:19}M.Yu. Khlopov and A.G. Polnarev, Phys.Lett.B. 97 (1980), 383.
\bibitem{rf:20} M.Yu. Khlopov, B.E. Malomed and Ya.B. Zeldovich, MNRAS 215 (1985), 575.
\bibitem{rf:21}P.D. Naselsky and A.G. Polnarev, Sov.Astron. 29 (1985), 487.
\bibitem{rf:22}B.J. Carr and J.E. Lidsey, Phys.Rev.D. 48 (1993), 543.
\bibitem{rf:23} B.J. Carr, J.H.Gilbert and J.E. Lidsey, Phys.Rev.D. 50 (1994), 4853.
\bibitem{rf:24}P. Ivanov, P. Naselsky and I. Novikov, Phys.Rev.D. 50 (1994), 7173.
\bibitem{rf:25}J. Garcia-Bellido, A. Linde and D. Wands, Phys.Rev.D. 54 (1997), 6040.
\bibitem{rf:26}A.M. Green \& A.R. Liddle, Phys.Rev.D. 56 (1997) 6166; 60 (1999), 0635616.
\bibitem{rf:27}L. Randall, M. Soljacic and A.H. Guth, Nuc. Phys. B. 472 (1996), 377.
\bibitem{rf:28}J. Yokoyama, Astron. Astrophys. 318 (1997), 673.
\bibitem{rf:29}J.S. Bullock and J.R. Primack, Phys.Rev.D. 55 (1997), 7423.
\bibitem{rf:30}P. Ivanov, Phys. Rev. D. 57 (1998), 7145.
\bibitem{rf:31}M. Crawford and D.N. Schramm, Nature 298 (1982), 538.
\bibitem{rf:32}S.W. Hawking, I. Moss and J. Stewart, Phys.Rev.D., 26 (1982), 2681.
\bibitem{rf:33}D. La and P.J.Steinhardt, Phys.Lett.B. 220 (1989), 375.
\bibitem{rf:34}A.G. Polnarev and R. Zemboricz, Phys.Rev.D. 43 (1988), 1106.
\bibitem{rf:35}S.W. Hawking, Phys. Lett. B. 231 (1989), 237.
\bibitem{rf:36}J. Garriga and M. Sakellariadou, Phys.Rev.D. 48 (1993), 2502.
\bibitem{rf:37}R. Caldwell and P. Casper, Phys. Rev.D. 53 (1996), 3002.
\bibitem{rf:38}J.H. MacGibbon, R.H. Brandenberger and U.F. Wichoski, Phys. Rev. D. 57 (1998), 2158.
\bibitem{rf:39}J.H. MacGibbon, Nature 329 (1987), 308.
\bibitem{rf:40}J.D. Barrow, E.J. Copeland and A.R. Liddle, Phys.Rev.D. 46 (1992), 645.
\bibitem{rf:41}K. Kohri and J. Yokoyama, preprint (1999); poster presentation.
\bibitem{rf:42}B.J. Carr, Ap.J. 206 (1976), 8.
\bibitem{rf:43}D.N.Page and S.W. Hawking,  Ap.J. 206 (1976), 1.
\bibitem{rf:44}J.H. MacGibbon and B.J. Carr,  Ap.J. 371 (1991), 447.
\bibitem{rf:45}J.H. MacGibbon, Phys. Rev.D. 44 (1991), 376.
\bibitem{rf:46}J.H. MacGibbon and B.R. Webber, Phys.Rev.D. 41 (1990), 3052.
\bibitem{rf:47} E.L. Wright, Ap.J. 459 (1996), 487.
\bibitem{rf:48}K. Maki, T. Mitsui and S. Orito, Phys.Rev.Lett. 76 (1996), 3474.
\bibitem{rf:49}D.B.Cline, D.A. Sanders and W. Hong, Ap.J. 486 (1997), 169.
\bibitem{rf:50}B.J. Carr, {\it Black Holes and High Energy Astrophysics} (Universal Academy Press, 1998), p 315. 
\bibitem{rf:51}K. Jedamzik, Phys.Rev.D. 55 (1997), R5871; Phys. Rep. 307 (1998), 155.  
\bibitem{rf:52}K. Jedamzik and J. Niemeyer  Phys.Rev.D. 59 (1999), 124014.
\bibitem{rf:53}T. Nakamura, M. Sasaki, T. Tanaka amd K. Thorn, Astrophys.J. 487 (1997), L139.
\bibitem{rf:54}K. Ioka, T. Tanaka and T. Nakamura  et al., Phys. Rev. D. 60 (1999), 083512.
\bibitem{rf:55}K. Ioka, Astrophys.J. (1999) (in press).
\bibitem{rf:56}T. Kaluza, Sitz.d.Preuss.Akad.d.Wiss.Physik-Mat.Klasse (1921), 966.
\bibitem{rf:57}P.A.M.Dirac, Proc. Roy. Soc. A165 (1938), 199.
\bibitem{rf:58}P. Jordan, {\it Schwerkraft und Weltall}, Vieweg (Braunschweig, 1955).
\bibitem{rf:59}D. Sciama, MNRAS 113 (1957), 34.
\bibitem{rf:60} C. Brans and R. H. Dicke, Phys. Rev. 124 (1961), 925. 
\bibitem{rf:61}C.M. Will, {\it Theory and experiment in gravitational physics} (Cambridge, 1993). 
\bibitem{rf:62}P.G. Bergmann, Int.J.Theor.Phys. 1 (1968), 25.
\bibitem{rf:63}R.V.Wagoner, Phys.Rev.D.1 (1970), 3209.
\bibitem{rf:64}K. Nordtvedt, Astrophys. J. 161 (1970), 1059,
\bibitem{rf:65}A. Guth, Phys. Rev. D. 23 (1981), 347.
\bibitem{rf:66}P.G.O.Freund, Nuc. Phys. B. 209 (1982), 146.
\bibitem{rf:67}K. Maeda, Class. Quant. Grav. 3 (1986), 233.
\bibitem{rf:68}E.W. Kolb, M.J.Perry and T.P.Walker, Phys.Rev.D. 33 (1986), 869.
\bibitem{rf:69}C.G. Callan et al., Nuc. Phys. B. 262 (1985) 597.
\bibitem{rf:70}T. Damour and A.M. Polyakov, Nuc.Phys.B 423 (1994) 532.
\bibitem{rf:71}P.J. Steinhardt and F.S. Accetta, Phys. Rev. Lett. 64 (1990) 2740.
\bibitem{rf:72}J.D. Barrow and P. Parsons, Phys. Rev. D, 55, 1906 (1997).
\bibitem{rf:73} S. W. Hawking, Commun. Math. Phys. 25 (1972), 167.
\bibitem{rf:74} M. A. Scheel, S. L. Shapiro and S. A. Teukolsky, Phys. Rev. D. 51 (1995), 4208. 
\bibitem{rf:75} M. A. Scheel, S. L. Shapiro and S. A. Teukolsky, Phys. Rev. D. 51 (1995), 4236. 
\bibitem{rf:76} T. Harada, T, Chiba, K. Nakao and T. Nakamura, Phys. Rev. D. 55 (1997), 2024. 
\bibitem{rf:77} M. Campanelli \& C. O. Lousto, Int. J. Mod. Phys., D2, 451 (1993)
\bibitem{rf:78} K. A. Bronnikov, C. P. Constantinidis, R. L. Evangelista and J. C. Fabris, 
Preprint, gr-qc/9710092 (1997).
\bibitem{rf:79} H. Kim, Phys. Rev. D. 60 (1999), 024001.
\bibitem{rf:80} A. Einstein and E. G. Straus, Rev. Mod. Phys. 17 (1945), 120. 
\bibitem{rf:81} C. B. G. McIntosh, Phys. Lett. 43A(1973), 33.   
\bibitem{rf:82} J. Kerimo and D. Kalligas, Phys. Rev. D. 58 (1998) 104002.
\bibitem{rf:83} T. Jacobsen, Phys. Rev. Lett. 83 (1999), 2699.
\bibitem{rf:84} J. O'Hanlon and B. O. J. Tupper, Nuovo Cim. 7B (1972), 305.
\bibitem{rf:85} D. F. Torres, A. R. Liddle and F. E. Schunck, Phys. Rev. D. 57 (1998), 4821.
\bibitem{rf:86} D. F. Torres, Phys. Rev. D. 56 (1997), 3478. 
\bibitem{rf:87} A. Whinnett, Preprint, Class. Quant. Grav. 16 (1999), 2797.
\bibitem{rf:88} D. N. C. Lin, B. J. Carr and S. M. Fall, MNRAS 177 (1976), 51. 
\bibitem{rf:89} G. V. Bicknell and R. N. Henriksen, Ap. J. 219 (1978), 1043. 




\end{thebibliography}
\end{document}